\documentclass[prl,twocolumn,showpacs,amsmath,amssymb,superscriptaddress,floatfix]{revtex4-1}

\usepackage{graphicx}
\graphicspath{{figs/}}
\usepackage{dcolumn}
\usepackage{bm}
\usepackage{amssymb,mathtools}
\usepackage{booktabs}
\usepackage{siunitx}
\usepackage{color}
\usepackage{microtype}
\usepackage[T1]{fontenc}
\usepackage{lmodern} 
\usepackage{enumitem}

\usepackage{physics}
\usepackage{upgreek}

\allowdisplaybreaks
\usepackage{placeins}

\usepackage[colorlinks,plainpages=false,linkcolor=blue,urlcolor=blue,citecolor=blue,pdfpagemode=UseNone,pdfstartview=FitBH]{hyperref}

\usepackage{sidecap}

\renewcommand{\tablename}{Table}
\makeatletter\renewcommand{\fnum@table}[1]{\tablename~\thetable.}\makeatother

\makeatletter

\newcommand{\LNO}{La$_3$Ni$_2$O$_7$}
\newcommand{\dx}{$d_{x^2-y^2}$}
\newcommand{\dz}{$d_{3z^2-1}$}

\newcommand{\Lh}{\underline{L}}

\definecolor{citecolor}{rgb}{0.0,0.60,0.32}

\begin{document}

\title{Charge and spin instabilities in superconducting La$_3$Ni$_2$O$_7$}

\author{Xuejiao Chen}
\thanks{These authors contributed equally to this work.}
\affiliation{National Laboratory of Solid State Microstructures and Department of Physics, Nanjing University, Nanjing 210093, China}
\affiliation{Key Laboratory of Magnetic Materials and Devices and Zhejiang Province Key Laboratory of Magnetic Materials and Application Technology, Ningbo Institute of Materials Technology and Engineering, Chinese Academy of Sciences, Ningbo 315201, China\looseness=-1}

\author{Peiheng Jiang}
\thanks{These authors contributed equally to this work.}
\affiliation{Key Laboratory of Magnetic Materials and Devices and Zhejiang Province Key Laboratory of Magnetic Materials and Application Technology, Ningbo Institute of Materials Technology and Engineering, Chinese Academy of Sciences, Ningbo 315201, China\looseness=-1}

\author{Jie Li}
\affiliation{National Laboratory of Solid State Microstructures and Department of Physics, Nanjing University, Nanjing 210093, China}

\author{Zhicheng Zhong}
\email[]{zhong@nimte.ac.cn}
\affiliation{Key Laboratory of Magnetic Materials and Devices and Zhejiang Province Key Laboratory of Magnetic Materials and Application Technology, Ningbo Institute of Materials Technology and Engineering, Chinese Academy of Sciences, Ningbo 315201, China\looseness=-1}

\author{Yi Lu}
\email[]{yilu@nju.edu.cn}
\affiliation{National Laboratory of Solid State Microstructures and Department of Physics, Nanjing University, Nanjing 210093, China}
\affiliation{Collaborative Innovation Center of Advanced Microstructures, Nanjing University, Nanjing 210093, China}

\date{\today}

\begin{abstract}

Motivated by the recent discovery of superconductivity in La$_3$Ni$_2$O$_7$ under high pressure, we explore its potential charge and spin instabilities through combined model analysis and first-principles calculations. Taking into account the small charge-transfer nature of high valence nickel, a fully correlated two-cluster model identifies a lattice-coupled charge instability characterized by substantial short-range fluctuations of oxygen holes. This instability is corroborated by density-functional-theory plus $U$ calculations that also reveal a strong tendency towards concurrent antiferromagnetic ordering. The charge, spin, and associated lattice instabilities are significantly suppressed with increasing external pressure, contributing to the emergence of superconductivity in pressurized La$_3$Ni$_2$O$_7$. Carrier doping is found to effectively suppress these instabilities, suggesting a viable strategy to stabilize a superconducting phase under ambient pressure.

\end{abstract}

\maketitle

\emph{Introduction.}
The electronic phase diagram of cuprate high-temperature superconductors exhibits a plethora of complex ordering phenomena characterized by comparable energy scales or ordering temperatures~\cite{Keimer2015}. These intertwined, often competing orders arise from strong electronic correlations and encompass various symmetry-breaking orders such as charge and spin density waves. Unraveling the complex interplay between these orders and superconductivity is pivotal for deepening our understanding of unconventional superconductors and correlated materials more broadly, and remains a central theme in ongoing experimental and theoretical studies~\cite{Fradkin2015,Tranquada1995,Ghiringhelli2012,Comin2016,Frano2020}.

The recent discovery of superconductivity in the infinite-layer~\cite{Li2019,Li2020a,Zeng2020,Osada2020a,Osada2020b,Zeng2022} and quintuple-layer~\cite{Pan2022} Ruddlesden-Popper nickelates $R_{n+1}$Ni$_n$O$_{2n+2}$ ($R=$rare earth) marks a significant advancement in the study of unconventional superconductors. These compounds, derived from the parent series $R_{n+1}$Ni$_n$O$_{3n+1}$ via chemical reduction by removing apical oxygens, feature stacked NiO$_2$ planes with Ni of electronic configuration close to 3$d^9$ with a single partially occupied 3\dx{} orbital, akin to the isostructural and isoelectronic cuprates~\cite{Wu2020,Kitatani2020} with a few noted differences~\cite{Botana2020}. Indeed, early experiments have unveiled symmetry-breaking competing orders in these compounds reminiscent of those in the cuprates~\cite{Krieger2022,Tam2022,Rossi2022}.

\begin{figure*}[t]
  \includegraphics[width=\textwidth]{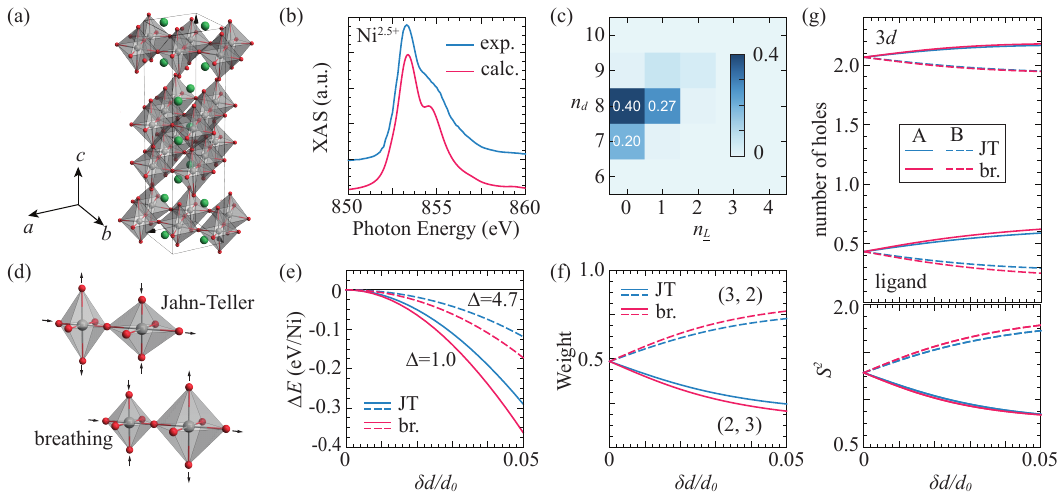}
  \caption{\label{fig:model} (a) Crystal structure of \LNO{} of $Amam$ symmetry~\cite{Sun2023}. (b) Experimental and model calculated Ni $L_3$-edge XAS for Ni$^{2.5+}$ with $\Delta=1.0$ eV. (c) Configuration weights of ground state projected on to a single NiO$_6$ octahedron. The weights of dominant configurations are listed. (d) The Jahn-Teller (JT) and breathing (br.) distortion patterns associated with possible charge density fluctuations within a nickel bilayer in \LNO{}. Only one layer is depicted.  (e) The change of ground state energy $\Delta E=E(\delta d/d_0) - E(0)$ as a function of the ratio of Ni-O bond length change $\delta d/d_0$. Results for $\Delta=4.7$ eV are shown for comparison. (f) Probability of hole distribution $(n_{\underline{A}}, n_{\underline{B}})$ over the compressed ($A$) and expanded ($B$) octahedra. (g) Distribution of holes on the 3$d$ and ligand orbitals (top), and total spin (bottom) as a function of $\delta d/d_0$. The solid and dashed lines correspond to octahedra $A$ and $B$, respectively.}
\end{figure*}

Recently, superconductivity with a high transition temperature of 80~K has also been reported in \LNO{} under high pressure~\cite{Sun2023}. \LNO{} distinguishes itself from other known nickelate superconductors due to its unique lattice and electronic structure. As a member compound of the series $R_{n+1}$Ni$_n$O$_{3n+1}$, \LNO{} retains intact bilayers of corner-sharing NiO$_6$ octahedra [Fig.~\ref{fig:model}(a)], resulting in a formal Ni configuration of 3$d^{7.5}$ that deviates significantly from its cuprate cousins, which necessitates a multi-orbital description of its possible superconducting mechanisms~\cite{Luo2023,Zhang2023,Yang2023,Lechermann2023,Sakakibara2023,Gu2023,Shen2023,Christiansson2023,Cao2024}. This substantial distinction in the relevant low-energy degrees of freedom naturally raises expectations of a distinct and potentially more complex phase diagram of \LNO{} than the cuprates, introducing an exciting new platform for further exploration of superconductivity and its interplay with the competing charge and spin orders.

Within this broad context, we investigate in this paper the charge and spin instabilities in \LNO{} using a combination of model analysis and first-principles density functional theory (DFT) calculations. We identify a lattice-coupled charge instability arising from short-range fluctuation of oxygen holes. This instability is manifested as long-range charge and concurrent antiferromagnetic orders in DFT plus Hubbard $U$ calculations, which could survive in the form of charge and spin-density waves in reality, giving rise to anomalies in transport measurements~\cite{Kobayashi1996,Greenblatt1997,Wu2001,Liu2022}. The charge, spin, and associated lattice orders are found to be significantly curbed under high pressure, consistent with experimental findings of pressure enhanced metallicity~\cite{Wu2001,Liu2022}. We further show that these instabilities can be effectively suppressed by carrier doping, suggesting a promising route to stabilize a superconducting phase under ambient pressure.

Before diving into the detailed calculations, we outline a rationale for the existence of the electronic and lattice instabilities in \LNO{}. The nominal Ni oxidation state in \LNO{} is $2.5+$, which can be viewed as a mixed valence state of $2+$ ($3d^8$) and $3+$ ($3d^7$). Trivalent Ni is relatively rare in oxides and is expected to have a small or negative charge-transfer energy in the Zaanen-Sawatzky-Allen classification scheme~\cite{ZSA}. Within this regime, the local configuration of the Ni$^{3+}$ ion approaches $3d^8$, where the additional electron is compensated by a hole on the surrounding oxygen ligands. The formal $3d^7$ configuration is then better represented by $3d^8\Lh$, where $\Lh$ denotes a ligand hole. Such a self-doping mechanism has been identified as crucial for understanding the electronic and magnetic properties of high valence cuprates~\cite{Mizokawa1991} and nickelates~\cite{Park2012,Johnston2014,Green2016,Lu2018}. In the case of \LNO{}, in a simplified picture, it is expected that on average each NiO$_6$ octahedron accommodates half a ligand hole, with ground state comprising primarily $3d^8$, $3d^8\Lh$ and $3d^7$ configurations. Such a dynamical valence fluctuation may couple to lattice distortions via electron-phonon coupling, resulting in charge order and lattice symmetry breaking at low temperatures~\cite{Goto2003}.

\emph{Charge instability in a two-cluster model.}
To explore the charge instability in \LNO{}, we construct a minimal model comprising two interconnected NiO$_6$ octahedra, each representing a sublattice site in a lattice with the simplest rocksalt-type distortion. Similar models have previously successfully examined charge and spin states in perovskite nickelates~\cite{Johnston2014,Green2016,Lu2018}. The Hamiltonian is given as
\begin{equation*}
  H = \sum_{\alpha,\alpha' \neq \alpha} H_d^\alpha + H_L^\alpha + \sqrt{1-\tau/2} V_{dL}^{\alpha\alpha} + \sqrt{\tau/2} V_{dL}^{\alpha \alpha'},
\end{equation*}
where $\alpha \in \{A, B\}$ labels different sublattices. $H^\alpha_L$ ($H^\alpha_d$) includes all local one-body (and Coulomb) interactions of the O-2$p$ (Ni-3$d$) orbitals, and $V_{dL}^{\alpha \alpha}$ ($V_{dL}^{\alpha \alpha'}$) defines the intra (inter) octahedra hybridization. $\tau$ is a semi-empirical parameter that interpolates between the limits of isolated ($\tau=0$) and dimerized ($\tau=1$) octahedra, while keeping the total Ni-O hybridization fixed. The details of model construction and parameter choices are described in the Supplemental Material~\cite{SupMat}. The model is solved using exact diagonalization as implemented in the \textsc{Quanty} code~\cite{Haverkort2012,Lu2014,Haverkort2016}.

To estimate the charge-transfer energy $\Delta$ and the value of $\tau$, we fit the calculated Ni $L_3$-edge (2$p_{3/2}\rightarrow 3d$) x-ray absorption (XAS) spectrum to the experimental one of Ni$^{2.5+}$, obtained by averaging the Ni$^{2+}$ and Ni$^{3+}$ spectra~\cite{Alders1998,Piamonteze2005}, which also agrees well with recent measurements on \LNO~\cite{Chen2024}. Calculation with $\Delta=1.0$ eV and $\tau=0.8$ well reproduces the experimental spectrum, as shown in Fig.~\ref{fig:model}(b). It is worth noting that the value of $\Delta$ falls between those estimated for Ni$^{2+}$ ($4.7$ eV)~\cite{Haverkort2012} and Ni$^{3+}$ ($-1.0 \sim -0.5$ eV)~\cite{Green2016,Lu2018}, reflecting the intermediate nature of the mixed valence state of \LNO. The weights of ground-state configurations with local occupation $(n_d, n_{\Lh})$ on a single octahedron are plotted in Fig.~\ref{fig:model}(c). As anticipated, the ground state consists of predominantly $d^8$ and $d^8\Lh$ configurations with weights 0.40 and 0.27, respectively, followed by $d^7$ with a weight of 0.20. Averaging over all configurations results in $\langle (n_d, n_{\Lh}) \rangle=(7.92,0.44)$, with approximately half a ligand hole per octahedron.

We now proceed to study the lattice coupled charge instability by considering two distortion patterns of the NiO$_6$ octahedra as sketched in Fig.~\ref{fig:model}(d). The Jahn-Teller (breathing) pattern is characterized by out-of-phase (in-phase) elongation and contraction of the in-plane and out-of-plane Ni-O bonds, both preserving the local tetragonal symmetry. For simplicity, each bond is assumed to be elongated or contracted with the same ratio $\pm \delta d/d_0$. More general distortion patterns with in-plane and out-of-plane anisotropy can be constructed from these two modes. For both modes, the energy decreases with increasing distortion $\delta d/d_0$ [Fig.~\ref{fig:model}(e)], reaching 0.3-0.4 eV/Ni for bond disproportionation at $\delta d/d_0=0.05$. This substantial energy gain could overcome the increase in lattice potential energy estimated in DFT calculations~\cite{SupMat}, realizing an ordered state. Notably, the lower lattice potential energy estimated for the Jahn-Teller distortion also suggests it as the dominant distortion type. We emphasize that the small charge-transfer energy is crucial here, as increasing it to that of NiO (4.7 eV) results in a reduction of the energy gain by more than 50\%, effectively suppressing the ordering tendency. Examination of the hole distribution $(n_{\underline{A}}, n_{\underline{B}})$ in Fig.~\ref{fig:model}(f), where $A$ ($B$) denotes the octahedron compressed (expanded) in plane, reveals that the holes move from the expanded octahedron to the compressed one. The redistribution of holes is more pronounced for the breathing distortion at a given value of $\delta d/d_0$, due to the more significant change of hybridization with simultaneous elongation or contraction of bond length~\cite{SupMat}.
The transferred holes in Fig.~\ref{fig:model}(f) is decomposed into the 3$d$ and ligand orbitals in the top panel of Fig.~\ref{fig:model}(g). The 3$d$ states remain close to a 3$d^8$ configuration with less pronounced redistribution of holes compared to the ligand staes due to the strong suppression of charge movement by the Coulomb interaction. The combined effect of the octahedral crystal field and Hund's coupling then results in fully filled $t_{2g}$ states and half-filled $e_g$ states with total spin $S=1$ for the 3$d$ shell, which is partially screened by antiferromagnetically coupled ligand holes. An effective spin $S^2_\mathrm{eff} \approx 1.4$ is expected from an equal mixing of $1/2$ ($d^8\Lh$) and $S=1$ ($d^8$) in the undistorted case, close to the calculated value of $1.3$ shown in the bottom panel of Fig.~\ref{fig:model}(g). In the extremely disproportionated case where the ligand hole resides fully on the compressed octahedron, total spin of each octahedron $S_A^2$ and $S_B^2$ is expected to converge to values of $3/4$ and $2$, respectively, which aligns with the calculation.

The above semi-quantitative model analysis clearly reveals  the existence of an electronic instability resulting from charge fluctuations that couple strongly to lattice distortions in \LNO{}. Whether such fluctuations eventually condense into a bond ordered, charge disproportionated, and/or magnetically ordered state at low temperatures depends on material details beyond the current model.

\begin{figure}[tb]
  \includegraphics[width=\columnwidth]{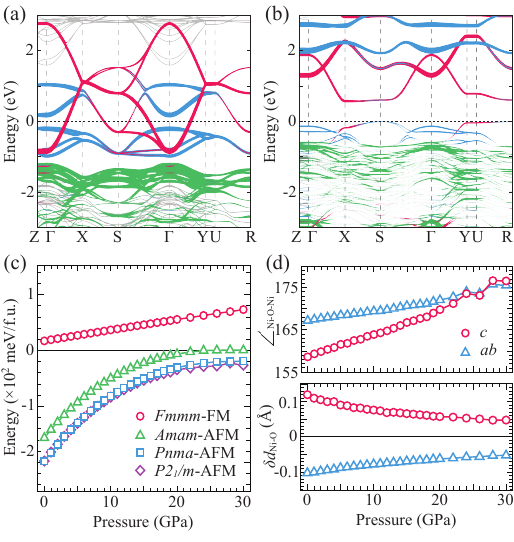}
  \caption{\label{fig:dft} DFT+$U$ band structure of \LNO{} for the $Amam$ nonmagnetic (a) and $Pnma$ $A$-type antiferromagnetic (b) cases. The width of red, blue, and green lines represents the weight of Ni \dx{}, \dz{}, and $t_{2g}$ characters, respectively. (c) DFT+$U$ total energy as a function of external pressure with the $A$-type AFM $Fmmm$ energy set as the reference point. For brevity, only $A$-type configurations are shown for AFM, which are the most stable ones at $U_\mathrm{eff}=4$ eV. (d) Pressure dependent Ni-O-Ni bond angles (top) and Ni-O bond lengths differences (bottom) for $Pnma$ with $A$-type AFM configuration.}
\end{figure}

\emph{Charge and spin order in DFT+$U$ calculation.}
To gain a more comprehensive and quantitative understanding of the ground state of \LNO{} with material specific details, we further perform DFT+$U$ calculations to determine the ground state crystal structure of \LNO{} and possible charge and spin orders. The calculations are performed using the Vienna \emph{Ab initio} Simulation Package (VASP)~\cite{vaspa,vaspb} with the Perdew-Burke-Ernzerhof (PBE) exchange-correlation functional~\cite{PBE}, with a $k$-point mesh of $9 \times 9 \times 3$ and an energy cutoff of 520 eV. The correlated Ni-3$d$ shell is treated on a mean-field level with an effective Coulomb interaction $U_\mathrm{eff}=U-J=4$ eV unless otherwise specified. This value is comparable to those commonly used for nickel oxides in similar studies~\cite{Dudarev1998,Park2012}. Additional calculations using a wide range of $U_\mathrm{eff}$ values up to 7 eV and the meta-generalized-gradient SCAN functional~\cite{SCAN} are also performed to assess the robustness and reliability of our findings~\cite{SupMat}. Assuming a rocksalt-type Ni-O bond-order distortion within each Ni bilayer, the experimental higher symmetry space groups $Amam$ and $Fmmm$ of \LNO{} under ambient and pressurized conditions~\cite{Sun2023} branch down to their subgroups $Pnma$ or $P2_1/m$ depending on the relative phase of distortion between neighboring bilayers. To find the most stable structure of \LNO{} under different pressures, the lattice parameters and internal atomic positions are relaxed using nonmagnetic (NM) as well as spin polarized calculations assuming ferromagnetic (FM) and different $A$, $C$, and $G$ types of antiferromagnetic (AFM) spin configurations within each space group~\cite{SupMat}.

In the nonmagnetic calculation, the $Amam$ structure is found to be lowest in energy at ambient pressure. The corresponding band structure, depicted in Fig.~\ref{fig:dft}(a), reveals fully filled $t_{2g}$ and partially filled $e_g$ states. The electron occupation projected onto the Ni-3$d$ Wannier basis is approximately 8.2, comparable to the value obtained from the above model analysis, indicating the existence of substantial ligand holes. The optimized $Pnma$ and $P2_1/m$ structure coincide with the $Amam$ one within numerical accuracy, showing no sign of bond order depicted in Fig.~\ref{fig:model}(d). The $Fmmm$ structure is higher in energy by about 20 meV per formula unit (f.u.). The energy difference gradually decreases with increasing pressure, and disappears at about 6 GPa, at which point the $Amam$ structure turns into $Fmmm$ with suppressed octahedral tilts, consistent with the results in Ref.~\cite{Sun2023}.
The situation changes drastically when the spin polarization is taken into account. With moderate $U_\mathrm{eff}$ values, the total energy decreases significantly by forming localized moments with long range orders~\cite{SupMat}. One example for the $A$-type antiferromagnetic configuration is shown in Fig.~\ref{fig:dft}(b). Compared to the FM state which remains metallic~\cite{Pardo2011}, AFM states exhibit a spin gap in the band structure and possess significantly lower energies, as shown in Fig.~\ref{fig:dft}(c). Under ambient pressure, comparison between the total energy of $Amam$ and $Pnma$ or $P2_1/m$ structures in Fig.~\ref{fig:dft}(c) shows that a concurrent bond and charge order further reduces the total energy by a sizable $\sim$ 50 meV/f.u., indicative of strong mutual feedback between the electronic and lattice degrees of freedom~\cite{Zhang2020stripe}. The two distorted structures show negligible energy difference, confirming a weak inter-bilayer coupling of the electronic structure. With increasing pressure, the bond-order energy gain gradually reduces to about 20 meV/f.u. at 30 GPa, suggesting a suppressed tendency towards structural distortion and associated charge order. We note that under high pressure, an antiferromagnetic stripe phase with a double-sized magnetic unit cell emerges as the most stable one with slightly lower energy compared to the A-type solution~\cite{Zhang2020stripe}, where a similar trend of pressure-suppressed bond order is also observed~\cite{SupMat}. Fig.~\ref{fig:dft}(d) shows the pressure dependent Ni-O-Ni bond angles and Ni-O bond length differences for the $Pnma$ structure with $A$-type antiferromagnetic configuration, both characterizing a gradually reduced structural distortion with increasing pressure. Here, $\delta d_\mathrm{Ni-O}$ is defined as the in ($ab$) and out-of-plane ($c$) Ni-O bond length differences between the two octahedra, and the different signs indicate a dominant Jahn-Teller type distortion. At 20 GPa, the octahedron-averaged $\delta d_\mathrm{Ni-O}$ is about $0.05$~\AA, corresponding to $\delta d/d_0=0.013$ in Fig.~\ref{fig:model} for calculated average $d_0=1.91$ \AA{}, which produces only a small charge and moment disproportionation. Similar results are obtained for $P2_1/m$~\cite{SupMat}. As it is known from the cuprates that the charge and spin instabilities compete with superconductivity~\cite{Keimer2015,Fradkin2015}, their suppression under pressure could be crucial for the emergence of superconductivity in \LNO{} in experiment~\cite{Sun2023}.

\begin{figure}[tb]
  \includegraphics[width=\columnwidth]{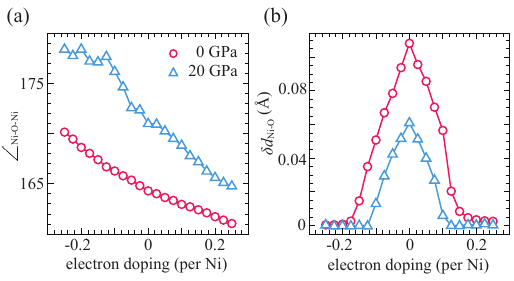}
  \caption{\label{fig:doping} Doping dependent Ni-O-Ni bond angles (a) and Ni-O bond length differences (b). The values are averaged over the in-plane and out-of-plane ones.}
\end{figure}

A natural question to follow is how these competing instabilities can be suppressed to facilitate superconductivity without resorting to high pressure conditions. One possible route is to introduce carrier doping. Fig.~\ref{fig:doping} shows the average Ni-O-Ni bond angles and Ni-O bond length differences across a wide effective doping range under 0 and 20 GPa by adjusting total electron number in the calculation. Under both pressures, the bond angles exhibit a monotonic decrease with increasing electron doping, indicating that doping serves as an effective means of controlling the correlated Ni-3$d$ bandwidth. More importantly, the bond order is quickly suppressed with increasing doping, suggesting that doping can effectively disrupt the bond disproportionation and potentially enhance the stability of the system against charge and spin density waves or ordering. A consistent trend of suppressed bond order upon doping is also realized by substituting La with divalent ions~\cite{SupMat}. Further experimental exploration of both the effects of doping on bandwidth control and the suppression of bond order, alongside strain engineering in thin films~\cite{Mochizuki2018}, holds the potential to stabilize a superconducting phase in \LNO{} under ambient pressure.

\emph{Summary and Discussion.}
In summary, we have studied the possible charge and spin instabilities in \LNO{} using a combination of model analysis and first-principles calculations. With realistic, material specific parameters, a minimal, fully correlated two-cluster model reveals substantial ligand holes in the ground state, as verified by recent spectroscopic studies~\cite{Dong2023,Chen2024}. The short-range fluctuation of these ligand holes results in a charge instability that couples strongly to lattice distortions. Such an instability is manifested as a long range static charge order with concurrent antiferromagnetic order and lattice distortions in DFT+$U$ calculations over a wide range of effective $U$ values. The charge, spin and associated lattice instabilities are found to be greatly suppressed with increasing external pressure, which could be one of the key factors contributing to the emergence of superconductivity in pressurized \LNO~\cite{Sun2023}. A small amount of carrier doping is found to be effective in suppressing these instabilities, which could be a promising route to stabilize a superconducting phase under ambient pressure.
We should emphasize, however, that DFT+$U$ treats the correlated problem on a mean-field level, and is known to overestimate a system's tendency towards ordering. The insulating, long-range ordered ground state found here could well be metallic with only short-range, incipient orders or fluctuations, as indicated by the experimental transport measurements~\cite{Kobayashi1996,Wu2001,Liu2022}. It is plausible to speculate that \LNO{} resides in close proximity to a quantum critical point associated with charge ordering and/or antiferromagnetic behavior. A closely related example is the perovsikte LaNiO$_3$, where the theoretically identified orders~\cite{Subedi2018} are found to be density waves of the same momentum space structure in experiments~\cite{Li2016,Zhang2017,Guo2018,Liu2020}. We thus expect the overall picture obtained here to be qualitatively correct, but the details of the charge and spin instabilities and their interplay with superconductivity in \LNO{} remain to be further explored in future studies. It is also worth noting that other compounds in the Ruddlesden-Popper series $R_{n+1}$Ni$_n$O$_{3n+1}$, where Ni is assigned a nominal valence of $(3-1/n)+$ and thus possess a small charge-transfer energy for $n \geq 2$, present a tantalizing possibility of harboring a superconducting phase frustrated by charge and/or spin density waves of similar nature~\cite{Zhang2020,Li2024,Zhu2024,Zhang2024}.

\begin{acknowledgments}
Y. L. acknowledges support from the National Key R\&D Program of China (No. 2022YFA1403000) and the National Natural Science Foundation of China (No. 12274207). Z. Z acknowledges support from the National Key R\&D Program of China (Nos. 2021YFA0718900 and 2022YFA1403000), the Key Research Program of Frontier Sciences of CAS (No. ZDBS-LY-SLH008), the National Nature Science Foundation of China (Nos. 11974365, 12004400 and 51931011), the K.C. Wong Education Foundation (No. GJTD-2020-11), and the Science Center of the National Science Foundation of China (No. 52088101).
\end{acknowledgments}

\bibliographystyle{apsrev4-1}
\bibliography{lno327_bo_abbrev.bib}
\clearpage
\onecolumngrid
\begin{center}
\large{\textbf{Supplemental Material for \\ ``Charge and spin instabilities in superconducting La$_3$Ni$_2$O$_7$''}}
\end{center}

\setcounter{figure}{0}
\setcounter{table}{0}
\setcounter{equation}{0}
\setcounter{page}{1}
\renewcommand{\thefigure}{S\arabic{figure}}
\renewcommand{\thetable}{S\Roman{table}}
\renewcommand{\theequation}{S\arabic{equation}}

\def\@hangfrom@section#1#2#3{\@hangfrom{#1#2}#3}
\def\@hangfroms@section#1#2{#1#2}

\section*{S1 -- Double-cluster model}

\subsection*{S1.1 -- Model construction and parameters}
The model consists of two interconnected NiO$_6$ clusters~\cite{Green2016,Lu2018}, with the full Hamiltonian given as
\begin{equation}
  H = \sum_{\alpha,\alpha' \neq \alpha} H_d^\alpha + H_L^\alpha + \sqrt{1-\tau/2} V_{dL}^{\alpha\alpha} + \sqrt{\tau/2} V_{dL}^{\alpha \alpha'}.
\end{equation}
where $\alpha \in \{A, B\}$ labels the two sublattice sites. The local term of the $d$ site reads
\begin{equation}
  H_d = \sum_{ij} \epsilon_d^{ij} d^\dagger_{i} d_{j} + \sum_{ijkl} U^{ijkl} d^\dagger_{i} d^\dagger_{j} d_{l} d_{k},
\end{equation}
where $d_i^{(\dag)}$ is the fermionic annihilation (creation) operator for the 3$d$ state with spin-orbital index $i$. For brevity, the site index $\alpha$ will be omitted hereafter when no confusion arises. The one-body energy tensor $\epsilon_d$ contains the on-site energies of the 3$d$ orbitals and an atomic spin-orbit coupling interaction. The two-body part contains the full Coulomb interaction tensor $U_{ijkl}$, parametrized using Slater integrals $F^k$ and $G^k$~\cite{Haverkort2012}. The local term for the ligand site is given as
\begin{equation}
  H_L = \sum_{ij} \epsilon_L^{ij} L^\dagger_{i} L_{j},
\end{equation}
where $L_i^{(\dag)}$ represents operators for the symmetry adapted ligand orbitals (i.e. molecular orbitals obtained from linear combination of O-2$p$ orbitals). The hybridization terms read
\begin{equation}
  V_{dL}^{\alpha \beta} = \sum_{i} t_{dL}^{i} ({d^\alpha}^\dagger_{i} L^\beta_{i} + \text{h.c.}).
\end{equation}
For the $D_4$ local symmetry, the $d$ and ligand orbitals transform as four irreducible representations (irreps) \{$a_{1g}$ ($3z^2-1$), $b_{1g}$ ($x^2-y^2$), $b_{2g}$ ($xy$), and $e_g$ ($xz/yz$)\}. (Note that in the main text, we group orbitals by the irreps of $O_h$ group for convenience, where $e_g$ includes $3z^2-1$ and $x^2-y^2$, and $t_{2g}$ includes $xy$, $xz$, and $yz$.)

The one-body parameters are calculated in the basis of maximally localized Ni-3$d$ and O-2$p$ Wannier orbitals~\cite{wien2k,Mostofi2014} obtained from DFT calculation of \LNO{} using a fully relaxed $Fmmm$ structure with no external pressure. The DFT derived on-site energies, in the order of irreps, are given as $\epsilon_d = E_d + \{ 0.41, 0.46, -0.36, -0.25 \}$ (in units of eV hereafter) for the 3$d$ orbitals and $\epsilon_L = E_L + \{ 0.52, 0.22, -0.36, -0.19\}$ for the ligand orbitals, and the corresponding hybridizations are $t_{dL} = \{2.33, 2.75, 1.53, 1.39\}$. Since the parameters predominantly depend on the local octahedral environment of Ni, values derived from both experimental~\cite{Sun2023} and DFT-calculated $Amam$ structures without pressure exhibit marginal differences owing to slight variations in bond lengths. The spin-orbit coupling for the 3$d$ states are also included with coupling constant $\lambda_d=0.1$. The mean on-site energies $E_d$ and $E_L$ are related via the charge transfer energy $\Delta$, which we define as the energy cost of transferring one electron from the fully filled ligand states to the $3d$ states, namely $\Delta = E(3d^{8}3d^{8}\Lh)-E(3d^{7}3d^{8})$. We have taken $U_{dd}=6.0$ and $\Delta=1.0$ for calculations presented in the main text. For spectroscopic calculations, additional Ni 2$p$ core states are also included, with core-valence interaction $U_{cd}=7.5$ and spin-orbit coupling $\lambda_c=11.5$. The multipole part of the valence-valence and core-valence Coulomb interaction parameters are taken as 80\% of the Hartree-Fock values~\cite{Cowan1981} of Ni$^{2+}$, which are $F_{dd}^2=12.23$, $F_{dd}^4=7.60$, and $F_{cd}^2=7.72$, $G_{cd}^1=5.78$, $G_{cd}^3=3.29$.

To simulate the effects of lattice distortion, we scale the hybridization parameters associated to a elongated/contracted bond following Harrison's rule~\cite{harrison2012}, where $t^{(e/c)}_{dp}/t^0_{dp}=(1\pm \delta d/d_0)^{-4}$. For the breathing distortion with isotropic bond length changes for a cluster, this means the same overall scaling of $t_{dL}$. For the Jahn-Teller distortion, the hoppoings associated with individual Ni-O bonds are scaled before transforming into $t_{dL}$ of molecular orbitals in different irreps.

To account for the reduced dimensions in the bilayer \LNO{}, we use an increased value of the inter-cluster hopping scaling factor $\sqrt{\tau/2}=0.40$ compared to the values $0.33$ or $0.35$ used in Refs.~\cite{Green2016,Lu2018}. This adjustment is consistent with the fact that an $A$($B$) octahedron in the bilayer structure is surrounded by 5 instead of 6 $B$($A$) octahedra in the perovsikte structure, which implies an approximate increase by a factor of $\sqrt{6/5} \approx 1.1$ for the inter-cluster hybridization.

\subsection*{S1.2 -- Estimation of lattice potential energy}
To estimate the increase of lattice potential energy in the bond ordered state, we calculate the energy increase in DFT when bond disproportionation is introduced to a tetragonal ($I4/mmm$) \LNO{} structure with optimized lattice constants and internal atomic positions with no external pressure. The calculation is non-spin-polarized without including the Coulomb $U$, to minimize changes in the band structure and associated changes of the electronic energy. The calculation is performed using the Wien2k code~\cite{wien2k} with the PBE functional and a $k$-point mesh of $8\times8\times8$.

\begin{figure}[htb]
  \includegraphics[width=0.5\textwidth]{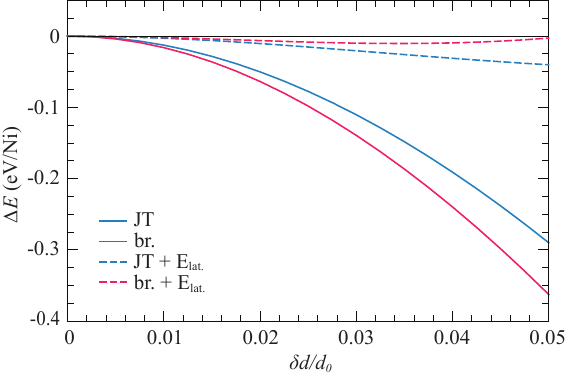}
  \caption{\label{Sfig:Elatt} Energy gain as a function of bond disproportionation $\delta d/d_0$, with (dashed) and without (solid) correction by the lattice potential energy $E_\mathrm{lat.}$.}
\end{figure}

Setting $\delta d/d_0$=0.04 results in an energy increase of 0.23 and 0.16 eV/Ni for the breathing and Jahn-Teller distortion, respectively. Assuming a quadratic dependence of the potential energy on $\delta d/d_0$, $E_{\mathrm{lat.}}$ is estimated as 144 $(\delta d/d_0)^2$ and 100 $(\delta d/d_0)^2$ eV/Ni, respectively. The corrected total energy is shown in Fig.~\ref{Sfig:Elatt}.

\subsection*{S1.3 -- Characterization of electronic structure with bond disproportionation}

\begin{figure}[htb]
  \includegraphics[width=0.5\textwidth]{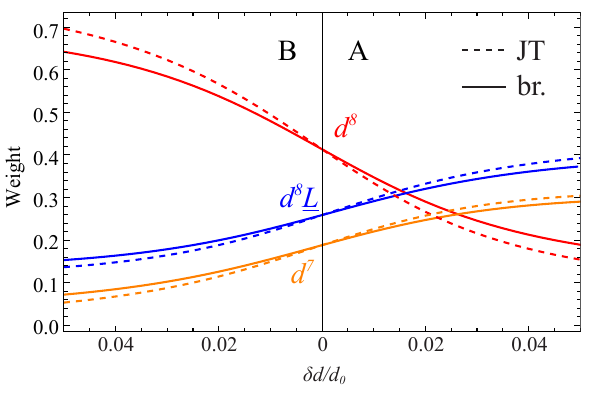}
  \caption{\label{Sfig:configs} Weight of the leading local configurations of the compressed $A$ and expanded $B$ NiO$_6$ octahedra as a function of bond disproportionation $\delta d/d_0$. The solid and dashed lines correspond to breathing and Jahn-Teller type distortions, respectively.}
\end{figure}

Figure~\ref{Sfig:configs} shows the dependence of local configurations of the compressed $A$ and expanded $B$ NiO$_6$ octahedra on bond disproportionation, which exhibits similarities for both types of distortion. As disproportionation increases, the $d^8$ configuration gradually localizes onto the expanded octahedra, while $d^8\Lh$ and $d^7$ show the opposite behavior.

\clearpage
\section*{S2 -- Spin configurations used in DFT+$U$ calculations}

\begin{figure}[htb]
  \includegraphics[width=0.75\textwidth]{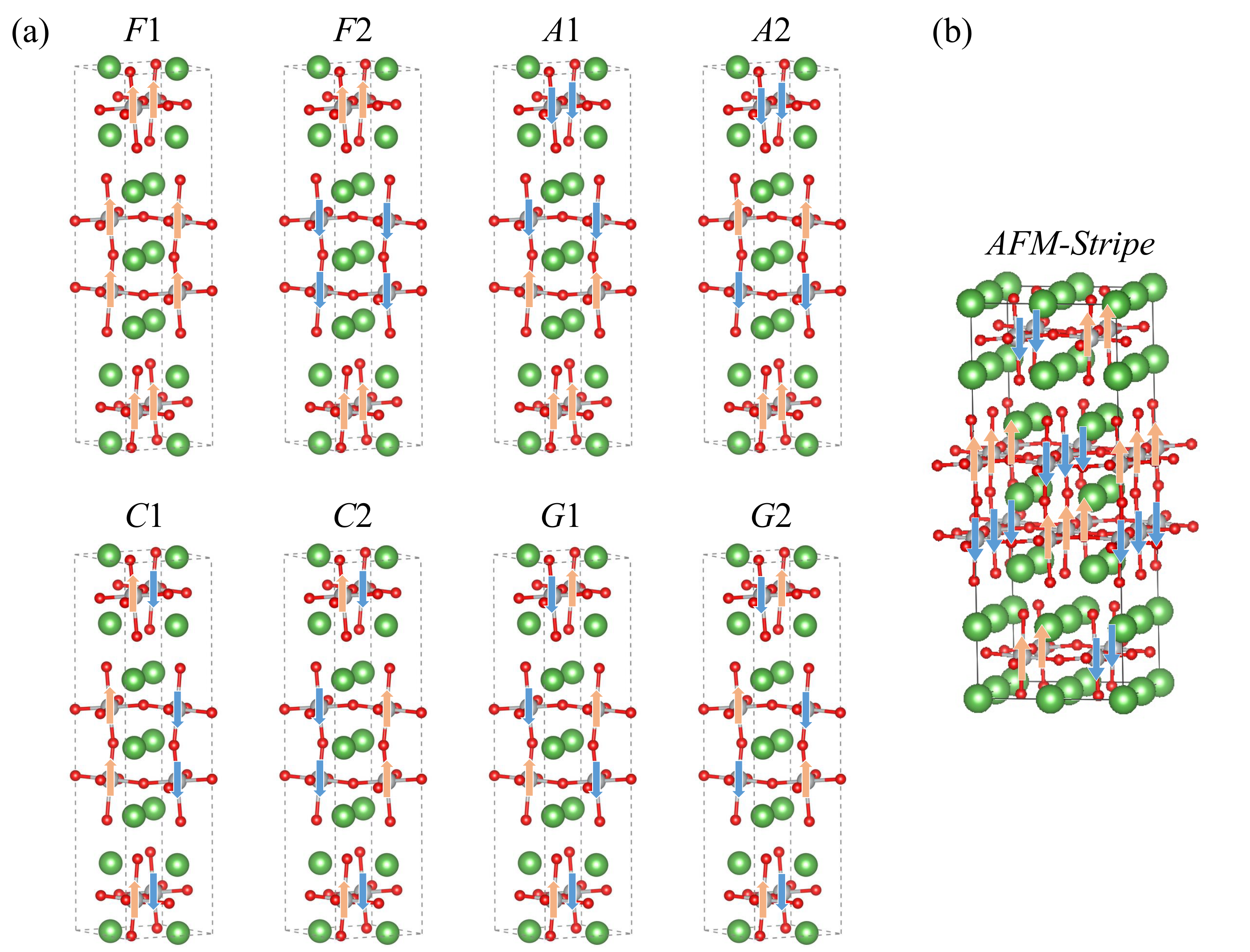}
  \caption{\label{Sfig:magconfig} Definition of ferromagnetic and antiferromagnetic spin configurations with a 4 f.u. unit cell (a) as well as an antiferromagnetic stripe order with a 8 f.u. unit cell (b). See also Ref.~\cite{Mochizuki2018}. }
\end{figure}

Fig.~\ref{Sfig:magconfig} summarizes the definition of magnetic orders considered in the DFT+$U$ calculations. Fig.~\ref{Sfig:magconfig}(a) shows the definition of ferromagnetic and antiferromagnetic spin configurations compatible with the 4 f.u. $Amam$, $Pnma$, or $P2_1/m$ unit cell. Fig.~\ref{Sfig:magconfig}(b) also shows an antiferromagnetic stripe order accommodated within a $\sqrt{2} \times \sqrt{2} \times 1$ unit cell, or $2 \times 2 \times 1$ of the pseudo-tetragonal unit cell. The ferromagnetic $F$ type and antiferromagnetic $A$, $C$, $G$ types follows the usual definition, and the numbers $1$ and $2$ labels two possible inter-bilayer arrangements. As the inter-bilayer coupling is found to be negligible in our calculations (on the order of a few meV, see Sec. S3), we omit the results of subtypes $2$ and refer to the $F1$,  $A1$, $C1$, and $G1$ types as $F$,  $A$, $C$, and $G$ in the main text.

\clearpage
\section*{S3 -- Unfolded DFT+$U$ band structures}
\begin{figure}[htb]
  \includegraphics[width=0.75\textwidth]{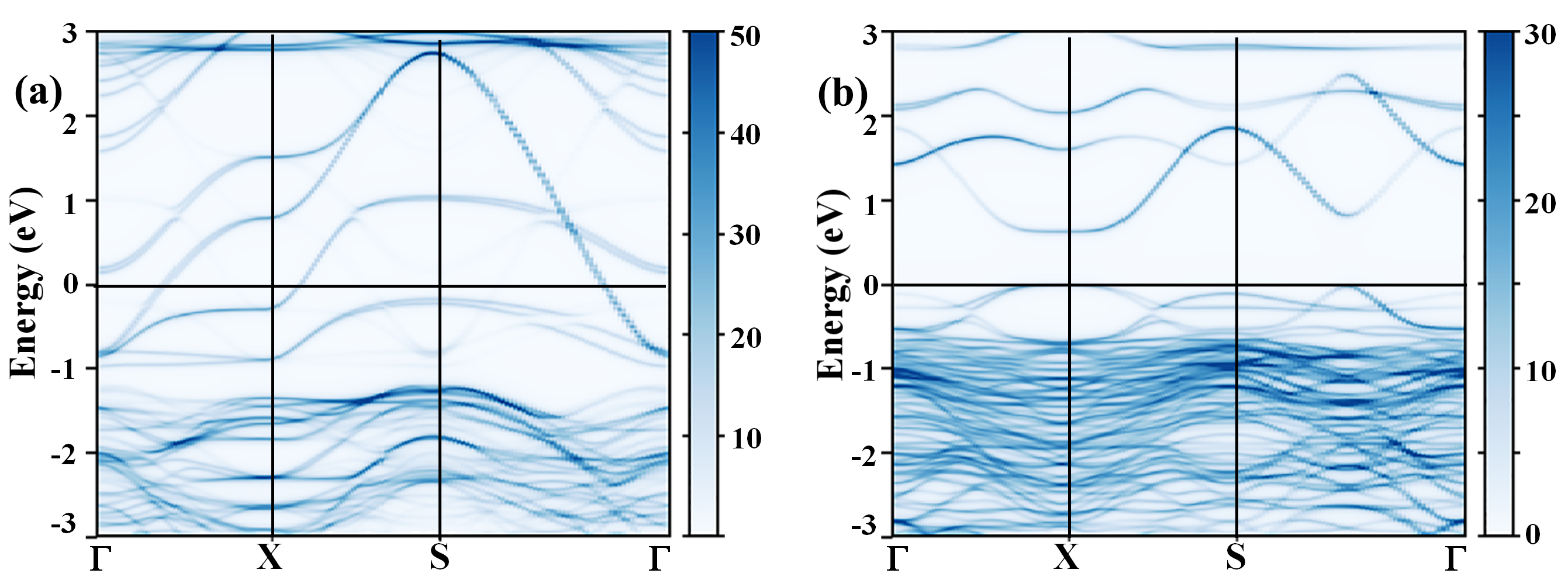}
  \caption{\label{Sfig:unfoldband} Unfolded DFT+$U$ ($U=$ 4.0 eV) band structure of \LNO{} for the $Amam$ nonmagnetic (a) and $Pnma$ A-type antiferromagnetic (b) cases. The color represents the weight (in percentage) of the spectral function.}
\end{figure}

Fig.~\ref{Sfig:unfoldband} shows the same DFT+$U$ band structures in Fig.~\ref{fig:dft} in the main text, unfolded onto the pseudo-tetragonal unit cell of \LNO{} along the high-symmetry $\mathbf{k}$-path $\Gamma(0,0)-X(\pi,0)-S(\pi,\pi)-\Gamma(0,0)$.

\clearpage
\section*{\label{Ssec:dftfull} S4 -- Additional DFT+$U$ results}

\subsection*{DFT+$U$ calculations with different $U$ values}

Table~\ref{Stab:totalE} summarizes the DFT+$U$ total energy (eV per 4 f.u.) calculated for different space groups under external pressures 0 GPa and 20 GPa for $U=$ 0, 2, 5, and 7 eV. The bond ordered phase $Pnma$ is always preferred with finite $U$ values. Note that the $P2_1/m$ phase is omitted in this wide parameter scan, as it is comparable to $Pnma$ in energy with negligible differences due to the weak inter-bilayer coupling of the electronic structure of \LNO{} (see main text). A general trend observed is that with increasing $U_\mathrm{eff}$ values, the most stable magnetic structures gradually evolve from $G$ to $A$, and possibly to $F$ types.

\begin{table}[htb]
  \renewcommand{\arraystretch}{1.5}
  \centering
  \begin{tabular}{ccccccccccc}
    \multicolumn{11}{c}{A -- 0 Gpa} \\ \toprule[1.0pt]
           &     SG & NM & F1 & F2 & A1 & A2 & C1 & C2 & G1 & G2 \\ \hline
           & $Amam$ & -380.1322 & -380.1308 & -380.1287 & -380.1318 & -380.1314 & -380.1291 & -380.1312 & -380.1323 & -380.1320 \\
       U=0 & $Pnma$ & -380.1319 & -380.1300 & -380.1288 & -380.1319 & -380.1310 & -380.1319 & -380.1299 & -380.1349 & -380.1281 \\
           & $Fmmm$ & -380.1035 & -380.1091 & -380.1021 & -380.1041 & -380.1044 & -380.1042 & -380.1041 & -380.1043 & -380.1043 \\
           \hline
           & $Amam$ & -371.3436 & -371.8035 & -371.8082 & -372.2463 & -372.2495 & -372.0936 & -372.0934 & -372.0513 & -372.0513 \\
       U=2 & $Pnma$ & -371.3445 & -371.8036 & -371.8083 & -372.3023 & -372.3038 & -372.0931 & -372.0919 & $\mathbf{-372.3141}$ & $\mathbf{-372.3116}$ \\
           & $Fmmm$ & -371.3423 & -371.4495 & -371.2108 & -371.2979 & -371.9875 & -371.3416 & -371.3417 & -371.3424 & -371.3421 \\
           \hline
           & $Amam$ & -359.3262 & -364.1598 & -364.1617 & -364.1691 & -364.1711 & -363.3684 & -363.3651 & -362.8723 & -362.8485 \\
       U=5 & $Pnma$ & -359.3279 & -364.1588 & -364.1612 & $\mathbf{-364.3384}$ & $\mathbf{-364.3398}$ & -363.6944 & -363.6945 & -362.2701 & -362.2656 \\
           & $Fmmm$ & -359.3288 & -363.9420 & -363.9442 & -359.3050 & -363.9598 & -359.3290 & -359.3278 & -359.3293 & -359.3286 \\
           \hline
           & $Amam$ & -352.2318 & $\mathbf{-360.4582}$ & $\mathbf{-360.4584}$ & -360.2958 & -360.2972 & -359.3781 & -359.3540 & -358.6881 & -358.6929 \\
       U=7 & $Pnma$ & -352.2311 & $\mathbf{-360.4582}$ & $\mathbf{-360.4580}$ & -360.4485 & -360.4494 & -358.3121 & -358.3272 & -359.2914 & -359.2911 \\
           & $Fmmm$ & -352.2327 & -360.3112 & -360.3102 & -351.7308 & -360.1079 & -355.9244 & -355.9222 & -352.2323 & -352.2321 \\
           \bottomrule[1.0pt]
  \\
    \multicolumn{11}{c}{B -- 20 Gpa} \\\toprule[1.0pt]
        &     SG & NM & F1 & F2 & A1 & A2 & C1 & C2 & G1 & G2 \\ \hline
        & $Amam$ & -311.5004 & -311.4995 & -311.4963 & -311.5002 & -311.5004 & -311.5000 & -311.4994 & -311.5009 & -311.5000 \\
    U=0 & $Pnma$ & -311.5004 & -311.4961 & -311.4966 & -311.5005 & -311.5004 & -311.5001 & -311.5006 & -311.5045 & -311.4970 \\
        & $Fmmm$ & -311.5000 & -311.5006 & -311.5049 & -311.5005 & -311.5007 & -311.5008 & -311.5008 & -311.5034 & -311.5003 \\
        \hline
        & $Amam$ & -303.1017 & -303.1392 & -303.1259 & -303.3471 & -303.3523 & -303.3979 & -303.3980 & $\mathbf{-303.4745}$ & $\mathbf{-303.4779}$ \\
    U=2 & $Pnma$ & -303.1015 & -303.1394 & -303.1405 & -303.3469 & -303.3514 & -303.3981 & -303.3978 & $\mathbf{-303.4771}$ & $\mathbf{-303.4778}$ \\
        & $Fmmm$ & -303.1013 & -303.1437 & -303.1446 & -303.3476 & -303.3529 & -303.1067 & -303.1064 & -303.1029 & -303.1011 \\
        \hline
        & $Amam$ & -291.6980 & -295.1036 & -295.1104 & -295.2645 & -295.2710 & -294.3392 & -294.6283 & -294.2933 & -294.1074 \\
    U=5 & $Pnma$ & -291.6986 & -295.1049 & -295.1100 & $\mathbf{-295.3949}$ & $\mathbf{-295.3921}$ & -294.6634 & -294.6638 & -295.3019 & -293.2709 \\
        & $Fmmm$ & -291.6985 & -295.1057 & -295.1141 & -295.2668 & -295.2725 & -293.7412 & -293.7440 & -291.7001 & -291.6985 \\
        \hline
        & $Amam$ & -285.0864 & -291.7102 & -291.7133 & -291.6139 & -291.6185 & -290.7926 & -290.7926 & -287.6204 & -287.5524 \\
    U=7 & $Pnma$ & -285.0751 & -291.7103 & -291.7122 & $\mathbf{-291.8147}$ & $\mathbf{-291.8754}$ & -289.9503 & -289.9470 & -289.7416 & -290.4238 \\
        & $Fmmm$ & -285.0872 & -291.7111 & -291.7140 & -291.6168 & -291.6185 & -287.9268 & -287.9268 & -285.0924 & -285.0869 \\
        \bottomrule[1.0pt]
  \end{tabular}
  \caption{\label{Stab:totalE} DFT+$U$ total energy with 0 Gpa (top) and 20 Gpa (bottom) pressure calculated with additional $U$ values from 0 to 7 eV. The lowest energy structure for each $U$ value is highlighted in bold.}
\end{table}

\subsection*{DFT total energy using SCAN functional}

Table~\ref{Stab:totalESCAN} summarizes the DFT total energy (eV per 4 f.u.) calculated for different space groups under external pressures 0 GPa and 20 GPa. Under 0 GPa pressure, similar to the DFT+$U$ results, $Pnma$ structure with $A$-type AFM configuration is the most stable one. Under 20 GPa, $Fmmm$ is found to be most stable, signifying a suppressed bond disproportionation, a trend in agreement with the DFT+$U$ results.

\begin{table}[htb]
  \renewcommand{\arraystretch}{1.5}
  \centering
  \begin{tabular}{cccccccccc}
    \toprule[1.0pt]
      Pressure (GPa)&     SG    & F1 & F2 & A1 & A2 & C1 & C2 & G1 & G2 \\ \hline
           & $Amam$ & -958.4288 &	-958.4290 &	-958.4520 &	-958.4517 &	-957.1955 &	-957.1910 &	-957.0992 &	-957.1017 \\
       0   & $Pnma$ & -958.4033	& -958.4026 &	$\mathbf{-958.5148}$ &	$\mathbf{-958.5156}$ &	-957.2325 &	-957.2338 &	-957.6177 &	-957.6163 \\
           & $Fmmm$ & -958.0313 &	-958.0329 &	-958.0635 &	-958.0664 &	-957.1157 &	-957.1159	& -956.8597 &	-956.8599 \\
           \hline
           & $Amam$ & -887.3327 &	-887.3381 &	-887.4669 &	-887.4710 &	-886.4859 &	-886.4851 &	-886.3801 &	-886.3807 \\
       20  & $Pnma$ & -887.3358 &	-887.3402 &	-887.4763	& -887.4804 &	-886.4999 &	-886.4997 &	-886.3974 &	-886.3959 \\
           & $Fmmm$ & -887.3038 &	-887.3094 &	$\mathbf{-887.4810}$ &	$\mathbf{-887.4858}$ &	-886.6021 &	-886.6019 &	-886.5052 &	-886.5052 \\
           \bottomrule[1.0pt]
  \end{tabular}
  \caption{\label{Stab:totalESCAN} DFT total energy with 0 Gpa (top) and 20 Gpa (bottom) pressure calculated with SCAN functional.}
\end{table}

\subsection*{Stripe orders in DFT$+U$ calculation}

To evaluate the robustness of the pressure dependence of the bond order and search for possible more complex magnetic states, we construct in-plane supercells ($\sqrt{2} \times \sqrt{2} \times 1$ of the $Amam$ unit cell, 8 f.u.) that accommodates antiferromagnetic stripe order with $q$=($\pi$,0) in the pseudo-tetragonal notation. Table~\ref{Stab:totalEStripe} summarizes the DFT+$U$ total energy (eV per 8 f.u.) calculated for different space groups under external pressures 0 GPa and 30 GPa for $U=$ 4 eV. For these supercell calculations we used a $k$-point mesh of $5 \times 5 \times 3$. Under 0 GPa pressure, $Pnma$ space group with A-type AFM configuration is the most stable one, followed by the stripe order at a slightly higher energy. At 30 GPa, the stripe order turns to be the magnetic ground state for all considered space group symmetries, consistent with findings in  Ref.~\cite{Zhang2023}. For either the AFM or stripe order, the $Pnma$ symmetry exhibits the lowest energy, indicating a strong tendency towards bond and charge ordering.


\begin{table}[htb]
  \renewcommand{\arraystretch}{1.5}
  \centering
  \begin{tabular}{cccc}
    \toprule[1.0pt]
      Pressure (GPa)&     SG    & A1 & AFM-Stripe  \\ \hline
           & $Amam$ &-733.0906  &-732.4854  \\
       0   & $Pnma$ & -733.4293  &-733.3624   \\
           & $Fmmm$ &-732.6023  &-732.6029  \\
           \hline
           & $Amam$ &-530.4129  & -530.8619 \\
       30  & $Pnma$ &-530.5883	& -531.1182 \\
           & $Fmmm$ &-530.4197	&-530.9495 \\
           \bottomrule[1.0pt]
  \end{tabular}
  \caption{\label{Stab:totalEStripe} DFT+$U$ ($U$= 4 eV) total energy of different symmetries and spin configurations with 0 Gpa (top) and 30 Gpa (bottom) pressure.}
\end{table}

\subsection*{Pressure dependence of $P2_1/m$ structure}

\begin{figure}[htb]
  \includegraphics[width=0.5\textwidth]{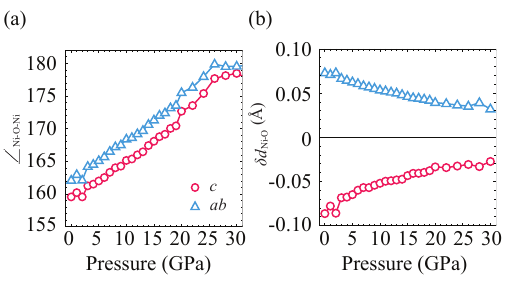}
  \caption{\label{Sfig:p21m} Pressure dependent Ni-O-Ni bond angles (a) and Ni-O bond lengths differences (b) for $P2_1/m$ with $A$-type AFM configuration.}
\end{figure}

Figure~\ref{Sfig:p21m} shows the pressure dependent Ni-O-Ni bond angles (top) and Ni-O bond lengths differences (bottom) for $P2_1/m$ with $A1$-type AFM configuration. A similar pressure dependence is observed for both the bond angles and bond disproportionation is observed as for the $Pnma$ phase (Fig.~\ref{fig:dft}) in the main text.

\clearpage
\section*{S5 - Effects of chemical doping on the structural properties}

\begin{figure}[htb]
  \includegraphics[width=0.75\textwidth]{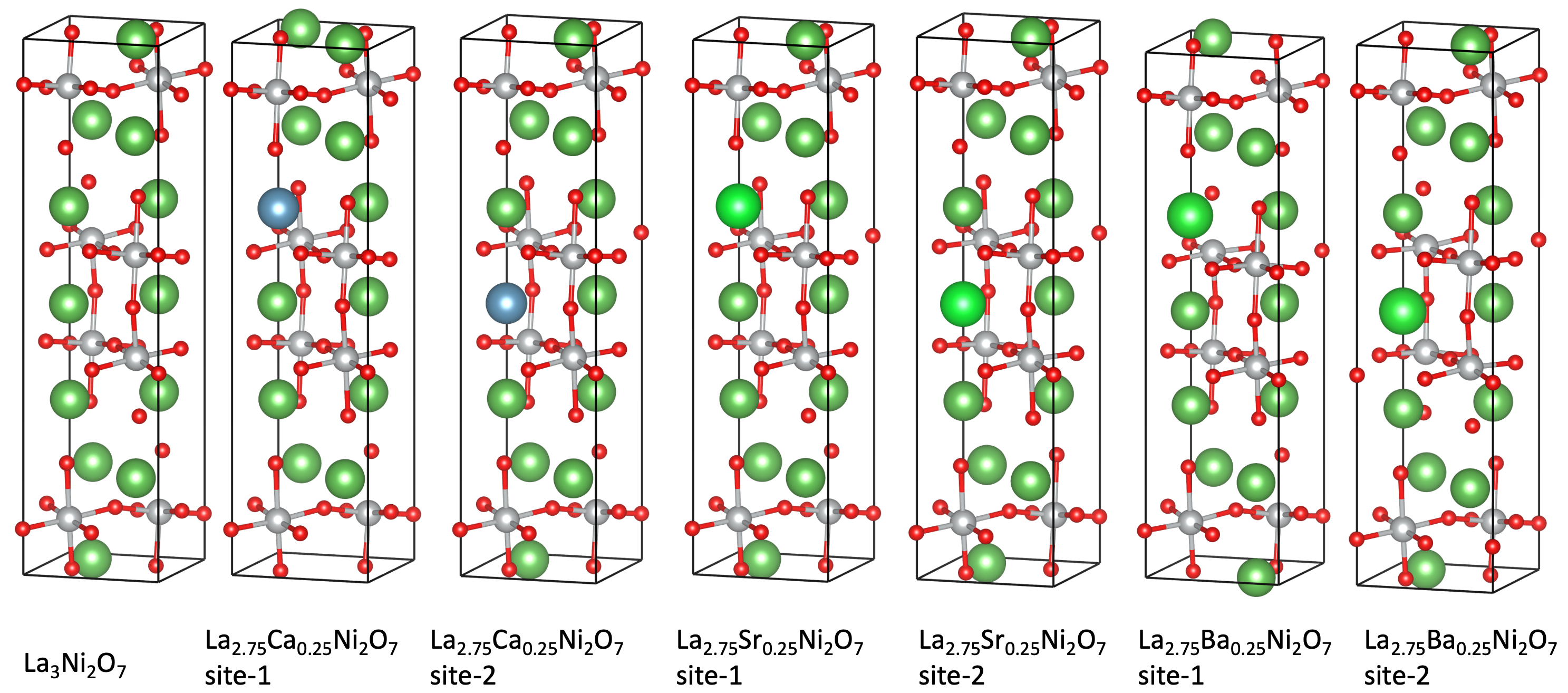}
  \caption{\label{Sfig:doping} Ca/Sr/Ba substitution of La at the outer (site-1) and inner site (site-2) of \LNO{} with $Pnma$ crystal structure.}
\end{figure}

We consider chemical doping by substituting La ions with divalent Ca, Sr, and Ba ions at two symmetrically distinct sites within a $Pnma$ unit cell of \LNO, as illustrated in Fig.~\ref{Sfig:doping}. This corresponds to a doping level of 0.125 hole/Ni. The structures are optimized within DFT+$U$ ($U$=4.0 eV) assuming AFM-A spin configuration, and the resulting bond disproportionation and Ni-O-Ni bond angles are summarized in Tab.~\ref{Stab:doping}. The reported values are averaged over all Ni sites within the unit cell. Overall, these substitutions tend to reduce bond length disproportionation, consistent with the findings of our simplified effective doping study in the main text. Moreover, an increase in the out-of-plane Ni-O-Ni bond angle is observed across all cases, which could increase the inter-bilayer antiferromagnetic superexchange coupling that has been proposed to be beneficial for superconductivity in recent theoretical studies~\cite{Cao2024}.

\begin{table}[htb]
  \renewcommand{\arraystretch}{1.5}
  \centering
  \begin{tabular}{lcccccccc}
    \toprule[1.0pt]
      Substitution & $\phantom{aaa}$ &     $|\delta d^{\mathrm{in}}|$    & $\phantom{aaa}$ & $|\delta d^{\mathrm{out}}|$ & $\phantom{aaa}$ & $\angle_{\mathrm{in}}$ & $\phantom{aaa}$ & $\angle_{\mathrm{out}}$  \\ \hline
       La$_3$Ni$_2$O$_7$            && 0.102 && 0.120 && 167.1 && 158.6 \\ \hline
       La$_{2.75}$Ca$_{0.25}$ (1)   && 0.099 && 0.023 && 165.3 && 159.8 \\
       La$_{2.75}$Ca$_{0.25}$ (2)   && 0.047 && 0.040 && 165.5 && 159.2 \\ \hline
       La$_{2.75}$Sr$_{0.25}$ (1)   && 0.008 && 0.009 && 166.1 && 160.6 \\
       La$_{2.75}$Sr$_{0.25}$ (2)   && 0.044 && 0.037 && 167.0 && 162.0   \\ \hline
       La$_{2.75}$Ba$_{0.25}$ (1)   && 0.055 && 0.031 && 166.0 && 160.6   \\
       La$_{2.75}$Ba$_{0.25}$ (2)   && 0.041 && 0.025 && 167.2 && 163.4   \\
    \bottomrule[1.0pt]
  \end{tabular}
  \caption{\label{Stab:doping} Structural effects of Ca/Sr/Ba substitution of La at the outer site (site-1) and inner site (site-2) of \LNO{} with $Pnma$ crystal structure.}
\end{table}

\end{document}